# Packaging-enhanced optical fiber-chip interconnect with enlarged grating coupler and multimode fiber

CHAO WANG,[1,2] CHINGWEN CHANG,[1] JASON MIDKIFF,[1] AREF ASGHARI,[1] JAMES FAN,[1] JIANYING ZHOU,[1] XIAOCHUAN XU,[1] HUIPING TIAN,[2*] AND RAY T. CHEN[1*]

[1]*Department of Electrical and Computer Engineering, The University of Texas at Austin, Austin, TX 78758, USA.*
[2]*State Key Laboratory of Information Photonics and Optical Communications, Beijing University of Posts and Telecommunications, Beijing 100876, China.*
*\*Corresponding author: chenrt@utexas.edu, hptian@bupt.edu.cn*

**Abstract:** Optical I/O plays a crucial role in the lifespan of lab-on-a-chip systems, from preliminary testing to operation in the target environment. However, due to the precise alignments required, efficient and reliable fiber to chip connections remain challenging, yielding inconsistent test results and unstable packaged performance. To overcome this issue, for use in single mode on-chip systems, we propose the incorporation of area-enlarged grating couplers working in conjunction with multimode fibers. This combination enables simpler, faster, and more reliable connections than the traditional small area grating coupler with single mode fiber. In this work, we experimentally demonstrate a 3dB in-plane (X, Y) spatial tolerance of (10.2 μm, 17.3 μm) for the large area configuration, being at least (2.49, 3.33) times that of the small area one, and agreeing well with theoretical calculations. The simple concept is readily applicable to a range of photonic systems where cheaper more robust optical I/O is desired.

## 1. Introduction

The rapid progress in photonic integration technologies is enabling optical applications in a wide range of areas, such as optical on-chip computing, communications, data conversion, sensing, etc. A chief technology of interest is that of optical I/O, playing a fundamental role in nearly all applications, as it comes into play across the entire life of a chip, from quality control in manufacturing to incorporation and operation in its target environment. In recent years, great efforts have been put into the exploration and development of highly efficient and reliable on-chip light coupling [1–6]. Additionally, light couplings with specific characteristics have been widely investigated, for example, those with polarization sensitivity [7–9], multi-band operation [10], wideband/broadband operation [11], custom directionality [12–15], multi-core integration [16], etc. In all of these works, 2D-vertical grating couplers (GCs) have been markedly exploited, taking advantage of their large freedom for on-chip location and integration. Nonetheless, in the majority of these works one common and critical issue continues today. This is the susceptibility to mode mismatch caused by fiber-chip misalignment, especially for the most widely adopted single mode fiber-chip applications. The lack of alignment tolerance habitually degrades coupling efficiency and renders the alignment process unreliable. To date, several works have investigated ways to improve the readout tolerance/reliability. Improving the spectrum bandwidth of the GC [11] is one of the feasible ways to improve the testing tolerance in the view of spectrum stability. This approach, however, still relies on a strict fiber tip to chip location, and so still suffers from the positioning alignment challenge. This issue is also why optical on-chip testing generally requires expensive high-precision positioning stages and highly trained professional operators.

Fortunately, efforts focused directly on improving the positioning tolerance have also been proposed and demonstrated, providing another feasible way to improve the readout reliability. The work in ref [17] demonstrated low loss couplings of 0.5dB and 1.1dB of the fundamental mode in a tapered waveguide from the single mode end (5×5 μm^2) to the multimode end (100×100 μm ^2), and from the mutimode end to single mode end, respectively. A low loss but more tolerant end-fire packaging could therefore be envisioned by the use of a larger beam size emitted from the multimode end of the waveguide taper. Based on this idea, the work in ref [18] went a step further and implemented the concept with an 8-channel wavelength division demultiplexer. Working with a 62.5 μm-diameter multimode fiber array, a mean insertion loss of 1.9dB was achieved. Further simulations showed that, assuming a uniform energy distribution of the receiving fiber and the same size input spot, a 1dB image shift tolerance of ±10 μm was possible. Later, the concept was widely adopted in various end-coupling and on-chip mode conversion applications. Specifically, for on-chip optical testing based on vertical light injection and emission, with the use of a mode conversion taper, the dimensions of the grating couplers can be designed to accommodate the dimensions of the injecting/receiving fibers. Apart from the aforementioned GCs in refs [1-16] with dimensions comparable to those of a SMF, few works have reported any further improvement. The work in ref [19] is one such existing work. By further enlarging the physical size of the grating coupler, the authors reported up to $\pm 5$ μm in-plane spatial tolerance with a 4.4dB insertion loss on a thin lithium niobate wafer. For a general standard of 3dB tolerance, the value is expected to be much smaller. However, a positioning tolerance at the several-micrometer level remains a critical challenge. An exceptional optical support system including precision positioners and highly skilled operators are always required. Thus, in consideration of the practical operations for on-chip testing and packaging, the reported couplings are still too position fragile. Moreover, in moving towards applications in miniaturized, handheld, or portable optical devices/systems, the coupling sensitivity to vibration is generally unacceptable. Therefore, further efforts are desired to facilitate quick and accurate testing as well as robust operation in the final product.

In this work, aimed at enlarging the mode distribution for more spatially tolerant light coupling, we propose the use of large area 2D subwavelength grating couplers (LGCs) in

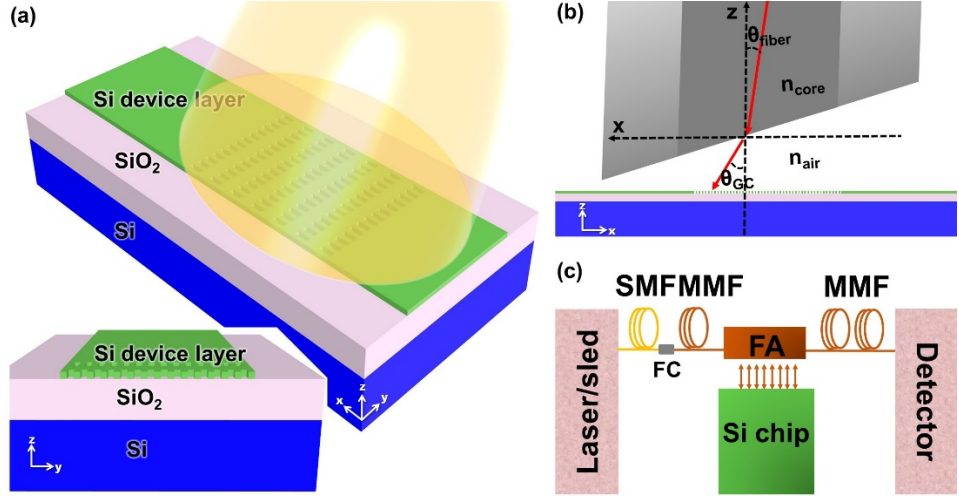

**Fig. 1** (a), schematic of the fiber-chip coupling; (b), schematic of the fiber-to-chip coupling in side view; (c), the optical path of the proposed MMF-LGC-MMF configuration; MMF: multimode fiber, SMF: single mode fiber, FC: fiber connector, FA: fiber array.

conjunction with multimode fiber (MMFs). Our LGCs were designed for operation at a wavelength of 1550 nm in the silicon-on-insulator (SOI) platform (as shown in Fig. 1(a)) using the same design method as in refs [20,21]. The LGCs were designed to work with 50 μm-diameter 8-degree end-surface polished MMFs. The optimized peak efficiencies for input (from the fiber to GC) and output (from the GC to air) were around 52% and 61%, respectively. The in-plane spatial tolerance of the fiber-chip alignment was analyzed both through simulation and experiment. In the simulations, only the fundamental mode is considered as the input activation mode and theoretically analyzed, as it's the working mode with the most energy transmitted/propagated in the silicon chip waveguides. The excited fiber mode is also primarily the fundamental mode. Following the most general standard, the 3dB spatial range of the in-plane fiber-chip positions (X, Y) was used for the spatial tolerance evaluations. The coordinates in Fig. (a) and (b) are defined the same, and figure (b) shows the schematic of fiber-to-GC coupling. In comparison to small area grating couplers (SGCs) with single mode fibers (SMFs), the proposed MMF-LGC-MMF solution (Fig.1(c)) improves the 3dB in-plane spatial tolerances from less than (4.1 μm, 5.2 μm) to (19.6 μm, 20.5 μm). Experimentally, 3dB (X, Y) spatial tolerances of (10.2 μm, 17.3 μm) were achieved. To the best of our knowledge this is the first time such a large on-chip light coupling spatial tolerance has been reported. This improvement eases the rigor of fiber-chip alignment and provides more stable signals in device operation.

## 2. Design and Theoretical Analysis

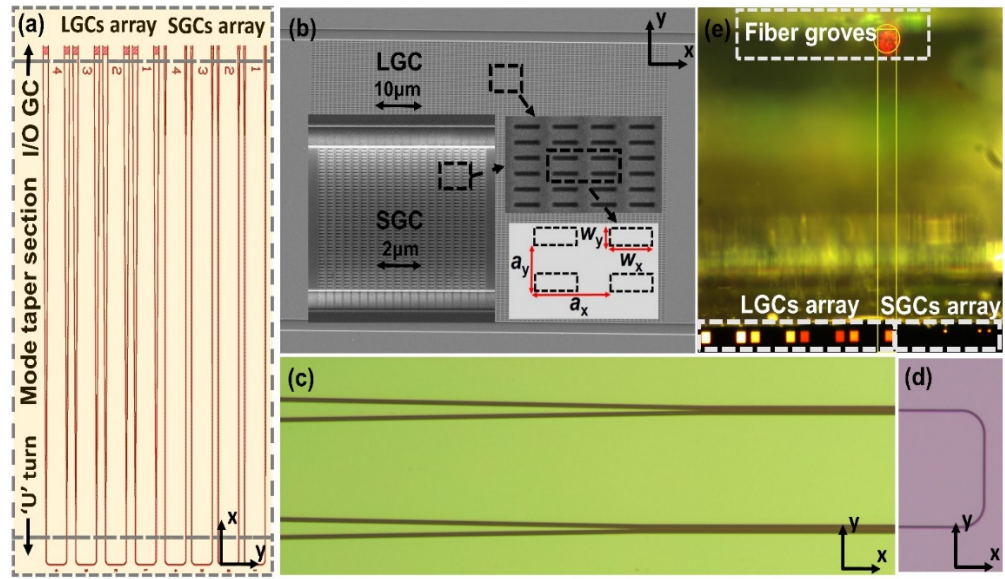

**Fig. 2.** (a), layout schematic of the fabricated chip; (b) SEM photos of the fabricated LGC and SGC. Insets: SEM photos of gratings and parameters. (c), (d), microscope images of the mode taper section the and single mode "U" turn silicon waveguide, respectively. (e), microscope picture of the fiber-chip coupling region, rough alignment assisted by red marker laser.

As shown in Fig. 2(a), the chip layout consisted of devices with SGCs or LGCs, straight/bend waveguides, and mode tapers. The SOI wafer consisted of a 13nm top silicon dioxide hard mask layer, a 220nm silicon layer, and a 3 μm buried oxide layer on a silicon substrate. Fig. 2(b) shows scanning electron microscopy (SEM) photos of fabricated GCs, with details of the gratings indicated. Two symmetric 3mm-long mode tapers (Fig. 2(c)) are connected to one another by a U-shape waveguide (Fig. 2(d)), and the on-chip circuits can be accessed by a fiber array (FA) (Fig. 2(e)). Apart from the difference in dimensions between the LGCs (length :

width = 80 μm:50 μm) and the SGCs (length : width = 20 μm:10 μm), all else was equivalent in their design and fabrication. And all the fibers used in this work were quartz fibers with the fiber tip polished with an angle of 8°. The diameters of the MMF and SMF are 50μm and 9μm, respectively. The LGC and SGC widths were set for mode matching to the MMF and SMF, respectively. Note that the LGC length of 80 μm is set to ensure that it covers the mode tail of the LGC output field. The insets of Fig. 2(b) show the details of the equivalent grating parameters for both GCs. The designed grating parameters were $(a_x, a_y, w_x, w_y)$ = (720, 380-410, 382, 96) nm, with post-fabrication devices possessing $(a_x, a_y, w_x, w_y)$ = (720, 390, 402, 123) nm. Note that a range of values of ay = 380, 390, 400, 410 nm was incorporated into the design to provide compensation for discrepancy between the designed and fabricated devices, which is practically inevitable and can adversely affect device performance.

The angle between the fiber and chip normal was adjusted to be about 10° for maximal output power. The corresponding working angle to/from the GC surface is given by:

$$\theta_{GC} = \sin^{-1}\left(\frac{n_{core}}{n_{air}} \sin \theta_{polish}\right) + \theta_{fiber} - \theta_{polish} \quad (1)$$

where $n_{core}$ = 1.44 and $n_{air}$ = 1 are the refractive indices of the fiber core and air, and $\theta_{polish}$ = 8° and $\theta_{fiber}$ = 10° are the fiber surface polished angle and fiber angle to the chip normal. With these values the working angle $\theta_{GC}$ is about 13.5°. We note that in the real world the coupling efficiency (CE) is affected by various setup factors (such as the fiber rotation and the fiber-to-chip separation). However, with the reasonable assumption that the measured power can be directly related to optimal power through a constant setup-induced loss factor α, i.e. $P_{3dB}$ = $(\alpha P)_{3dB}$, the various setup factors can be considered to have a negligible effect on the in-plane positioning tolerance evaluations. Thus, any discrepancy between experimental and theoretical powers is inconsequential in this work.

To theoretically evaluate the spatial tolerance of the fiber-to-chip coupling, we first determined the output modes from both the GC and the fiber—that is, the output mode from the GC activated by the TE-like fundamental mode of the connected silicon waveguide, and likewise the output mode from the MMF activated by the fundamental mode of the fiber. The mode profiles were calculated via a 3D-FDTD electromagnetic simulator (a commercially available software developed by Lumerical, Ansys Inc.). Then we performed the coupling analysis between the two modes using the Lumerical mode overlap analysis functions. We shifted the fiber in the X (parallel to the waveguide) and Y (perpendicular to the waveguide) directions and recorded the CEs. The in-plane directions (X, Y) were the focus of our analysis, as they are the predominant adjustments made in practical testing situations. We note again, as previously, that various other setup factors such as the fiber pitch angle, yaw rotation, and fiber-to-chip separation also affect the CE. However, these factors are typically easily controlled or pre-fixed, and thus are taken at their optimum values and not adjusted in the analysis. We did not consider light propagation in the chip, as it is time-prohibitive in 3D-FDTD simulations. Instead, we focused the coupling analysis on the far-field overlaps of the modes between the GC and the fiber where the coupling practically happens. Although this technique is not the most rigorous way to evaluate coupling performance, it's a sufficiently reliable way to quickly evaluate the effects of spatial alignment.

Coupling efficiency was calculated following [22]:

$$CE = \frac{P_{out}}{P_{in}} = Re\left[\frac{(\int \vec{E}_{out} \times \vec{H}_{in}^* \cdot d\vec{S}) \cdot (\int \vec{E}_{in} \times \vec{H}_{out}^* \cdot d\vec{S})}{\int \vec{E}_{out} \times \vec{H}_{out}^* \cdot d\vec{S}}\right] \frac{1}{Re(\int \vec{E}_{in} \times \vec{H}_{in}^* \cdot d\vec{S})}$$

(2)

where $(\vec{E}_{in}, \vec{H}_{in})$ and $(\vec{E}_{out}, \vec{H}_{out})$ are the far-field modes from either the GC or the fiber, depending on the direction of light propagation. Specifically, for setups with "fiber in," $(\vec{E}_{in},$ $\vec{H}_{in}$) is the far-field mode originating from the fiber fundamental mode, while ($\vec{E}_{out}, \vec{H}_{out}$) is the reflected far-field mode from the GC at the same location as ($\vec{E}_{in}, \vec{H}_{in}$). We note that the reflected mode could be used in the calculation because it carries the information of the mode coupled into the GC. On the other hand, for setups with "GC in," ($\vec{E}_{in}, \vec{H}_{in}$) is the far-field mode output from the GC activated by the fundamental mode of the connected on-chip silicon waveguide, while ($\vec{E}_{out}, \vec{H}_{out}$) is the far-field mode originating from the fiber fundamental mode at the same location. The calculation was carried out for deviations in both the X and Y directions, for four paths of interest: SMF-to-SGC, SGC-to-SMF, MMF-to-LGC, and LGC-to-MMF. Figures 3(a) and (c) show the results of these calculations where, to facilitate comparison, each curve has been normalized to its maximal value and takes the absolute value (to offset the effect on sign from reflection). From the plots we can extract the 3dB spatial tolerances (X, Y) of the CEs for each path: (9.1 μm, 8.0 μm), (4.6 μm, 6.8 μm), (28.6 μm, 28.1 μm), and (27.7 μm, 30.2 μm). We see improvements in the spatial tolerances of ~3–5 times for the large area configurations over the smaller ones. These improvements are direct consequences of the broadened mode sizes of the LGC and MMF, since each incremental spatial shift is a smaller percentage of the overall area, in comparison to the SGC and SMF.

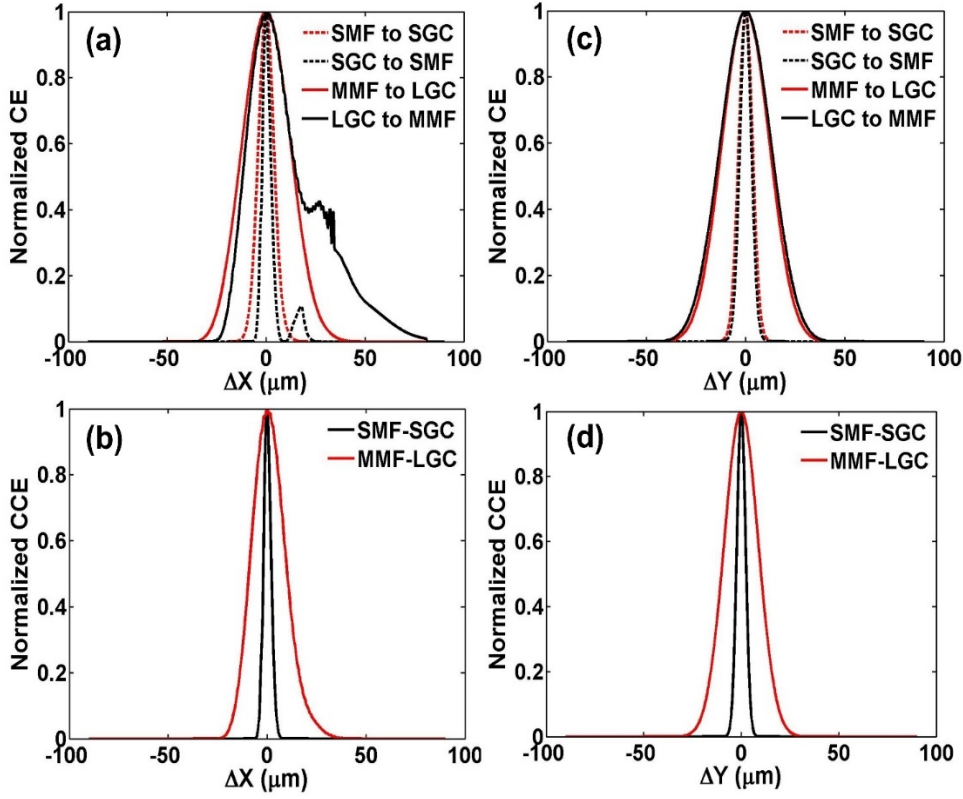

Fig. 3. (a) and (c) The calculated separate coupling efficiencies of the SMF to SGC, SGC to SMF, MMF to LGC, and LGC to MMF configurations versus X and Y deviations. (b) and (d) The comprehensive coupling efficiencies of the SMF to SGC to SMF, and MMF to LGC to MMF configurations versus X and Y deviations.

While the above calculations analyze the separate coupling of modes either from the fiber to the GC or from the GC to the fiber, in the interest of practical use, it is convenient to calculate the final coupling across both directions. That is to say, the "comprehensive coupling" from the fiber to the GC, and then back from the GC to the fiber. The comprehensive coupling efficiency (CCE) can be determined by multiplication of the component coupling efficiencies, that is:

$$CCE = \begin{bmatrix} CCE_x \\ CCE_y \end{bmatrix} = \begin{bmatrix} CE_{x12} * CE_{x21} \\ CE_{y12} * CE_{y21} \end{bmatrix} \quad (3)$$

where the value $CE_{x12}$ corresponds to the pre-calculated normalized coupling efficiency from fiber (1) to GC (2) with a deviation in the X direction, and similarly for the other values. Figures 3(b) and (d) show the normalized CCEs for X and Y deviations, respectively. Again, for comparison, calculations were made for both the large area and small area configurations, that is, for the now two paths of interest: SMF-to-SGC-to-SMF and MMF-to-LGC-to-MMF. (We note that, for the case of deviation in X, whereas side peaks were seen for the GC to fiber component CEs (Fig. 3(a)), these were suppressed and disappeared in the final CCE calculations (Fig. 3(b)). This point will be addressed in more detail in the follow-up experiment.) From these plots we can extract the final 3dB spatial tolerances (X, Y) of the CCEs for each path: (4.1 μm, 5.2 μm) and (19.6 μm, 20.5 μm). The spatial tolerance of the large area configuration is (4.78, 3.94) times that of the small area one. This improvement will help ease the alignment process and enable more reliable packaging, thus providing benefits to various optical applications.

## 3. Experiment

For experimental verification of our improved spatial tolerance concept, we fabricated and evaluated both the small area and large area GCs on the SOI platform, following our layout shown in Fig. 2(a). Each device structure consisted of two GCs (input and output) connected through silicon waveguide mode tapers to a central single mode "U-turn" waveguide. As previously described, grating parameters were varied across the devices to permit determination of the most optimal ones. The channel spacing was set to 250 μm to match the pitch of our fiber array. Chip fabrication was carried out with an electron-beam lithography one-layer exposure and a reactive-ion etch two-step process. For testing, the silicon chip was fixed by a vacuum aspirator on a chip holder, and the 8°-end-polished fiber array was adjusted to 10° with no horizontal rotation to match the GCs' working angle. For the initial rough alignment, a red light marker was injected through the fiber array to visually locate the input fibers and the GCs, as shown in Fig. 2(e). The hot rectangle spots seen in the figure are the GCs on the silicon chip. With this microscope view we note the greater visibility of the LGCs in comparison with the SGCs due to their larger physical size, providing the added benefit of simpler rough alignment for the user.

The fine (X, Y) positioning tolerance test was carried out by first searching for the optimal coupling position and taking this position as the center (0, 0). Then the fiber was shifted in 1 μm increments in the X or Y directions, letting the powers stabilize at the nanowatt level (the noise level) before taking a reading at each step. We mention that, while the actual positioning occurred in 1 μm increments, the resultant 3dB tolerances were extracted from fitted curves, providing a finer resolution to 0.1 μm. The fine alignment process was performed with a precision electronically controlled 4-degrees-of-freedom (X, Y, Z, R) positioning stage (SmarAct), where R is the horizontal rotation angle tuned to 0° and Z is the fiber-to-chip distance adjusted to be ~10 μm. Generally, because of the highly sensitive coupling versus the (X, Y) positions, a high-precision auto-controlled positioning stage is necessary for efficient, reliable, repeatable, and stable alignment in practical testing. The need for such tools is also one of the reasons why on-chip optical testing is so challenging and expensive. A wideband super-luminescent light-emitting diode (DenseLight CS5203A, 35mW) operating in the C band was used as the input source and a wideband power meter (Newport 1830-R-GPIB) was used as the power detector. We note that the experiment focused on the positioning tolerance evaluations and ignored analysis of the absolute output power, as it's affected by a variety of factors and has little influence on the tolerance evaluations. Figures 4(a) and (b) show the results of the measurements from the large area configuration for the X and Y deviations, respectively, along with the corresponding simulated data (the same curves as in Fig. 3(b) and (d)). Again, the measured powers have been normalized to their maximal values for ease of comparison. (We mention that the CCEs were also measured for the small area configurations, but that due to fragile coupling only 5 data points were collected for both the X and Y directions, being insufficient for curve fitting and therefore omitted from Fig. 4.) From the plots we extract the measured 3dB spatial ranges in the X and Y directions as 10.2 μm and 17.3 μm, respectively, agreeing well with the simulated data. Table 1 lists the details of the calculated and measured 3dB spatial tolerances for clear comparison. Taking into consideration all the configurations analyzed in Fig. 3 and Fig. 4, we determine that the large area configuration contributes at least (2.49, 3.33) times to the improvement of the spatial tolerances in the (X, Y) directions (the measured MMF-LGC-MMF result divided by the calculated SMF-SGC-SMF result).

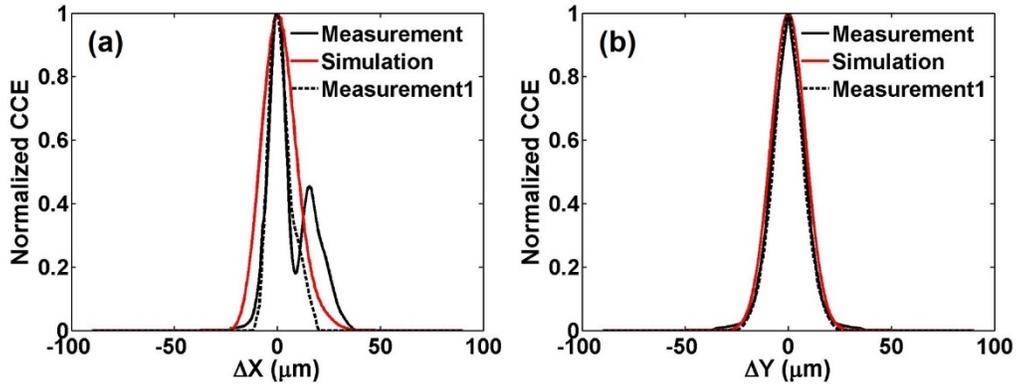

**Fig. 4.** (a), (b) Comparisons of experimental measurements and simulations of the comprehensive coupling efficiencies results X and Y deviations. The dark dashed curves are the additional measurements with shortened LGC length of 60μm.

| Table 1. Comparison of tolerances | | | |
|---|---|---|---|
| Setup | | 3dB spatial tolerance (μm) | |
| | | X | Y |
| Cal. | SMF-SGC | 9.1 | 8.0 |
| | SGC-SMF | 4.6 | 6.8 |
| | MMF-LGC | 28.6 | 28.1 |
| | LGC-MMF | 27.7 | 30.2 |
| | SMF-SGC-SMF | 4.1 | 5.2 |
| | MMF-LGC-MMF | 19.6 | 20.5 |
| Meas. | SMF-SGC-SMF | <4.1 | <5.2 |
| | MMF-LGC-MMF | 10.2 | 17.3 |

Apart from the main result, returning to Fig. 4(a) we acknowledge the presence of a side peak (or "fake peak" as we called it in testing) which appeared in the process of moving the fiber across the LGC. This peak is not consistent with the calculated plot of Fig. 3(b), suggesting some difference of significance between the simulation and experiment (e.g., only the fundamental mode of the fiber is considered in the simulations, while higher order modes may have some real effect [23]). We carried out several additional fabrications for further research on this issue and successfully removed the side peak without reducing the spatial tolerance

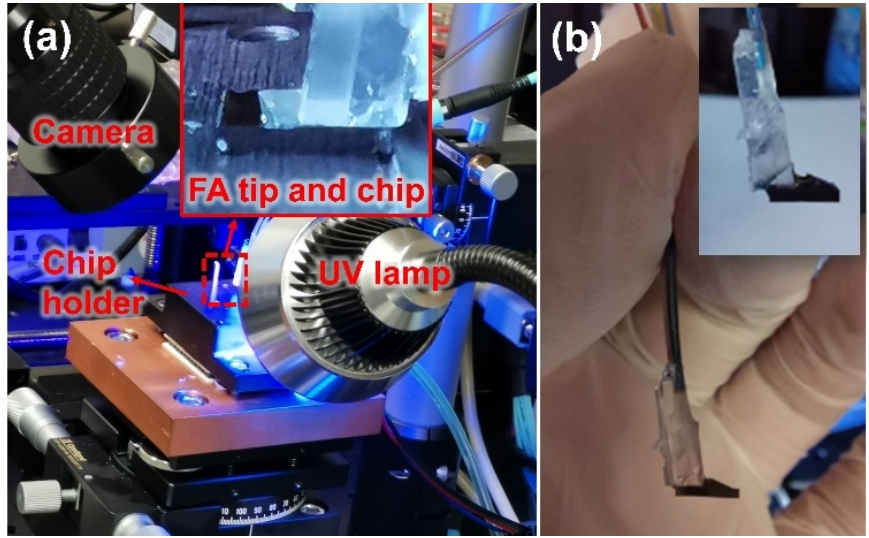

**Fig. 5** (a), setup for fiber-chip-glue testing; (b), the packaged fiber-chip module.

performance. Fig. 4(a) and (b) include results of an additional measurement on one such chip. The chip possessed the same design as the initial chips, except for a reduction of the LGC length from 80 μm to 60 μm. In fact, the length of the LGCs around the diameter of the fiber is considered sufficient (around 50 μm in this work), as the mode from the GC is primarily located near to the input waveguide connected to the GC.

To demonstrate the packaging reliability of the proposed LGC chip, we also conducted the fiber-to-chip attachment process, by means of a UV cure glue (Norland Optical Adhesive 61) and a 15W UV lamp. As shown in Fig. 5(a), the aligned chip (with an initial maximal output power of about 36 μW) was first mounted by negative pressure to the chip holder. Then, with the UV lamp turned on, 1–2 drops of UV cure glue were manually released by a needle tubing to the edge of the fiber chip. The output power was monitored in real-time throughout the whole process. In the beginning, it suffered a sudden slight drop, due to the impact of the glue drop, but recovered in a matter of seconds. It then decreased slowly during the exposure process. After about 10 minutes of UV exposure, the power stabilized. With an additional 20 minutes for enhanced curing, the attachment process was complete, with a final stabilized output power of 28 μW. We note that the reduction of balanced power was mainly due to the losses induced by the glue (absorption, scattering, and changed GC working angle). Figure 5(b) shows the assembled fiber-chip module. We emphasize that the attachment process was smooth and reliable, owing to the wide alignment tolerance provided by our large area coupling configuration. Much to the contrary, for the same operation with a small area coupler, the operation would be delicate and uncertain, due to the sensitivity of the coupling to physical vibration.

### 4. Further work and Conclusions

We point out that this work is an important first step in the demonstration of our proposed large area coupling concept. We have confirmed the enhanced spatial tolerance with basic mode tapers and waveguide paths. However, for increased confidence, it is necessary to evaluate the concept with more advanced device structures, such as optical modulators and sensors. Furthermore, to enable compatibility of our large area configuration with setups utilizing SMF input detectors, an additional connection to a fiber mode taper is required. For an optimized lensed fiber taper from a 62.5 μm end to a nanoend, an experimental loss as low as 0.013dB of the fundamental mode has been reported [24]. In fact, similar high transmission efficiency or efficiency optimization has been demonstrated with various fiber tapers in other works [25–

| Table 2. Mode conversion loss vs linear SOI waveguide taper ||||
| Multimode end (Height×Width) | Diameter of matched fiber | Saturated length | Minimal loss |
| --- | --- | --- | --- |
| 220nm×9µm | 9µm | 500µm | 0.032dB |
| 220nm×50µm | 50µm | 3000µm | 0.615dB |
| 220nm×62.5µm | 62.5µm | 5000µm | 0.895dB |
| 220nm×100µm | 100µm | 10000µm | 1.704dB |

28]. Thus, there will be only a slight and acceptable penalty in mode conversion loss incurred for the improvement in spatial tolerance acquired when combining the MMF of our configuration with an SMF input detector. Additionally, there will be tradeoffs between the conversion loss and the length of the on-chip SOI tapers. With an eigenmode expansion solver (Ansys Lumerical MODE) we analyzed the mode reshaping loss of a series of simple linear SOI tapers. The widths of the multimode ends were matched to the diameters of four types of multimode fibers, while the width of the single mode end remained constant at 500nm. The results are listed in Table 2. The saturated length is defined as the length at which the loss remains stable with further increase in length. We see that the larger the multimode end is, the longer the saturated length is, and the larger the minimal loss incurred, without other optimizations. The 50 μm diameter fiber was chosen as the basis of our work, considering the balance between chip size and acceptable loss.

In conclusion, recognizing the need for more practical and reliable fiber-to-chip optical interconnects, we have proposed an enlarged grating coupler to multimode fiber configuration, with the aim of improving the tolerance of the coupling efficiency to position. Comparing our large area configuration with that of the traditional small area one, theoretical analyses showed an improvement in the (X, Y) spatial tolerance of (4.78, 3.94) times. For experimental verification, we fabricated the grating couplers in the SOI platform and carried out a positioning analysis, demonstrating an improved (X, Y) spatial tolerance of at least (2.49, 3.33) times. To verify the concept further, we carried out a fiber-to-chip attachment process with real-time power monitoring. The process with our large area configuration was simple to execute and exhibited little power fluctuation, in comparison to typical experience with a small area configuration. Due to the straightforward working principle of our configuration—that is, the exploitation of a larger mode size for increased spatial tolerance—the concept is expected to be compatible with numerous optical on-chip devices and circuits. Through simpler fiber-chip alignment and more reliable packaging, the concept will lead to reduced operation costs and an accelerated development of integrated photonics for real world applications.

**Funding:**
The research was supported in part by the NIH Grant # 1 R43 AA026122-01 and in part by the National Science Foundation (NSF) under Contract 1932753. The authors also acknowledge the use of Texas Nanofabrication Facilities supported by the NSF NNCI Award #1542159. Chao's work is also supported by the National Natural Science Foundation of China (NSFC) (61634006, 61372038, 61431003); National Key Technologies R&D Program (2017YFA0205903, 2016YFB0402405); and the China Scholarship Council (CSC) (No. 201806470008).

**Disclosures:**
The authors declare no conflicts of interest.